\documentclass[reprint,aps,amsmath,amssymb,superscriptaddress,prd,nofootinbib]{revtex4-1}
 \pdfoutput=1

\def\tg{\tilde g}

\def\L{{\cal L}}

\newcommand{\eq}[1]{\eqref{#1}}
\newcommand{\be}[1]{\begin{equation}\label{#1}}
\newcommand{\ee}{\end{equation}}
\newcommand{\pd}[2]{\frac{\partial #1}{\partial #2}}

\usepackage{mathtools}
\newcommand{\F}{\prescript{*}{}{\cal{F}}} 
\newcommand{\A}{\prescript{*}{}{\cal{A}}} 

\begin{document}
\preprint{}
\title{Gravitational domain walls and the dynamics of $G$}


\author{Claudio Bunster}
\email{bunster@cecs.cl}
\affiliation{Centro de Estudios Cient\'{\i}ficos (CECs), Casilla 1469, Valdivia, Chile}

\author{Andr\'es Gomberoff}
\email{andres.gomberoff@uai.cl}
\affiliation{Facultad de Ingenier\'ia y Ciencias, Universidad Adolfo Ib\'a\~nez, Avda.~Diagonal las Torres 2640, Pe\~nalol\'en, Santiago, Chile}
\affiliation{Centro de Estudios Cient\'{\i}ficos (CECs), Casilla 1469, Valdivia, Chile}

\date{\today}

\begin{abstract}
From the point of view of elementary particle physics the gravitational constant $G$ is extraordinarily small. This has led to ask whether it could have decayed to its present value from an initial one commensurate with microscopical units. A mechanism that leads to such a decay is proposed herein. It is based on assuming that $G$ may take different values within regions of the universe separated by a novel kind of domain wall, a ``G-wall". The idea is implemented by introducing a gauge potential $A_{\mu\nu\rho}$, and its conjugate $D$, which determines the value of $G$ as an integration constant rather than a fundamental constant. The value of $G$ jumps when one goes through a $G$-wall. The procedure extends one previously developed for the cosmological constant, but the generalization is far from straightforward: (i) The intrinsic geometry of a $G$-wall is not the same as seen from its two sides, because the second law of black hole thermodynamics mandates that the jump in $G$ must cause a discontinuity in the scale of length. (ii) The size of the decay step in $G$ is controlled by a function $G(D)$ which may be chosen so as to diminish the value of $G$ towards the asymptote $G=0$, without fine tuning. It is shown that: (i) The dynamics of the gravitational field with $G$ treated as a dynamical variable, coupled to $G$-walls and matter, follows from an action principle, which is given. (ii) A particle that impinges on a $G$-wall may be refracted or reflected. (iii) The various forces between two particles change when a $G$-wall is inserted in between them. (iv) $G$-walls may be nucleated trough tunneling and thermal effects. The semiclassical probabilities are evaluated. (v)~If the action principle is constructed properly, the entropy of a black hole increases when the value of the gravitational constant is changed through the absorption of a G-wall by the hole.
\end{abstract}


\pacs{}

\maketitle


\section{Introduction}

A natural scale for the gravitational constant $G$ from particle physics would be set by the square of the inverse mass of the Higgs boson, $1/m_h^2\sim 10^{-4}GeV^{-2}$. The observed value of $G$ is  thirty four orders of magnitude smaller. This enormous difference led  Dirac to propose its ``large number hypothesis''\cite{Dirac:1937ti}, where $G$ was considered to change in time, decaying from an initial value conmensurable with microscopical units, to its value in the present epoch of the universe.  In this paper we propose a mechanism that leads to such a decay. It is based on assuming that $G$ may take different values within regions of the universe separated by a domain wall, which we call ``$G$-wall'' for short. The idea is implemented by introducing  a gauge potential $A_{\mu\nu\rho}$, and its conjugate $D$,  which determines the value of $G$ as an integration  constant rather than a fundamental constant. The value of the integration constant jumps when one goes through  a $G$-wall. 
 
The procedure extends one previously developed for the cosmological constant\cite{Brown:1988kg,Henneaux:1989zc,Gomberoff:2003zh}, but  the generalization is far from straightforward: (i) The intrinsic geometry of a $G$-wall is not the same as seen from its two sides. This is because the second law of black hole thermodynamics mandates that the jump in $G$ must cause a discontinuity in the scale of length. (ii) The size of the decay step in $G$ is controlled by a function $G(D)$ which may be chosen so as to diminish the value of $G$ towards the asymptote $G=0$, without fine tuning. 

The fact that $G$ sets the scale of length is similar to the way the effective gravitational constant changes in  the Jordan-Brans-Dicke theory\cite{Jordan:1949zz,PhysRev.124.925} or other dilatonic theories; but with the key difference that here the changes in $G$ are driven by the introduction of a $3$-form $A$, and its conjugate $D$, which do not have any local degrees of freedom.  

The plan of the paper is the following. Section II is devoted to introduce the key properties of the $G$-wall that separates regions of space with different values of $G$. The basis for the discussion is the introduction of the concept of ``gravitational units",  that will be key throughout the paper, and which incorporates the fact that the scale of length is changed when one crosses the $G$-wall. Two effects are displayed, namely, (i)the forces (gravitational and non-gravitational)  between two particles change in a distinct manner when a $G$-wall is inserted in between them,(ii) if a particle impinges on a $G$-wall, its worldline is refracted or reflected depending on its velocity. Section III deals with the action principle. The action is given, the equations of motion are derived; and the simple case in which the matter is a uniform vacuum energy (cosmological constant of microscopical origin) is dealt with in detail as a preparation for the study of $G$-wall nucleation in section IV. In that section, the nucleation through tunneling and thermal activation is discussed and the geometry and probability of the corresponding instanton and thermalon are studied. Finally, in section V, it is shown that the entropy of a black hole increases when it absorbs a $G$-wall. It is argued that the second law of black hole thermodynamics actually dictates the form of the action.
\section{Anticipation: Imprint of a $G$-wall}

Before tackling the general action principle, we will anticipate in this section, on the basis of a simple Newtonian argument, two effects which capture the distinctive imprint of a $G$-wall. They are: (a) the alteration in the forces (gravitational and non-gravitational)  between two particles when a $G$-wall is inserted in between them,  and (b) the effect of the wall on the motion of a test particle that crosses it.

\subsection{Effect of a $G$-wall on the forces between particles. Gravitational units}
\label{imprint}

Consider two particles interacting gravitationally through the Newtonian potential. The action is given by
\be{grava}
I=\int dt \left(\frac{m_1}{2}\vec v_1^2 + \frac{m_2}{2}\vec v_2^2 + \frac{m_1m_2G}{r_{12}} \right),
\ee
where $r_{12}=|\vec r_2-\vec r_1|$ is the distance between the particles. 

Next, perform a change of scale of the three fundamental units; time, length and mass,
\be{rescaling}
t=G^{1/2} \tilde t, \ \ \ \ \vec r=G^{1/2} \tilde{\vec r}, \ \ \ \ m=G^{-1/2}  \tilde m.
\ee
This change sets $G$ equals to unity, keeping the other two universal constants $\hbar$ and $c$ unaffected, as it leaves the units of action and speed invariant. We call these ``gravitational units''. This same rescaling was discussed, for example, in \cite{PhysRev.125.2163} in the context of the Jordan-Brans-Dicke theory\cite{Jordan:1949zz, PhysRev.124.925}.

In gravitational units the action \eq{grava} reads,
\be{grava2}
I=\int d\tilde t \left(\frac{\tilde m_1}{2}\vec v_1^2 + \frac{\tilde m_2}{2}\vec v_2^2 + \frac{\tilde m_1\tilde m_2}{{\tilde r}_{12}} \right).
\ee
The transition from Eq. \eq{grava} to Eq. \eq{grava2} was achieved assuming that $G$ was constant throughout space and time. We will now postulate that \eq{grava2} remains valid when $G$-walls are present, in which case $G$ is only domain-wise constant, since it changes its value when one crosses a $G$-wall. That is, we demand that { \em in gravitational units the $G$-wall becomes invisible in the action}.

\begin{figure}
\begin{center}
  \includegraphics[width=8cm]{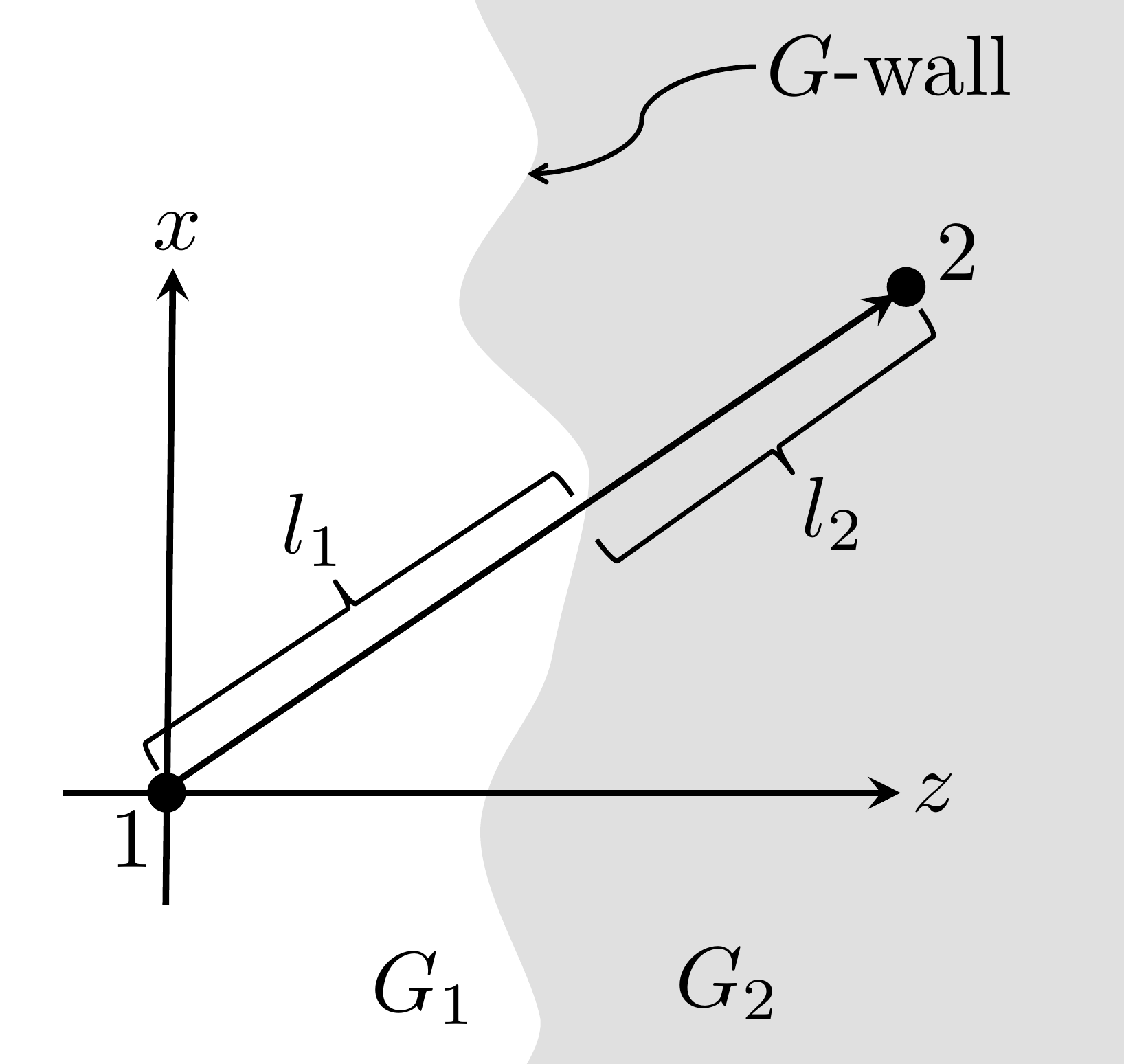}
  \caption{Two test particles, labeled 1 and 2, with  masses $m_1$, $m_2$ located at different sides of a $G$-wall, where the Newton constant is $G_1$, $G_2$ respectively. The  origin of the coordinate system is set on particle 1.}\label{newton}
\end{center}
\end{figure}

Consider now a $G$-wall as shown in Fig. \ref{newton}. Particle $1$, of mass $m_1$, is fixed at the origin of coordinates, to the left of the wall, while particle 2, of mass $m_2$, is located to its right. Let us first evaluate the gravitational force felt by particle 2 due to the field produced by particle 1. The potential energy is
$$
\tilde U=-\frac{\tilde m_1\tilde m_2}{{\tilde r_{12}}} \ ,
$$
where $\tilde r_{12} = \tilde l_1 + \tilde l_2$, is the 
sum of the tilded lengths of the portions, at different sides of the wall, of the line connecting particle 1 and particle 2. In terms of the original, untilded ``atomic units'' the potential energy of particle 2 in the field of particle 1 reads
\be{potG}
U_{1\rightarrow 2}= G^{-1/2}_2 \tilde U=   -\frac{\tilde m_1\tilde m_2}{\left(\sqrt{\frac{G_2}{G_1}}l_1 +l_2\right)} \ .
\ee
The force felt by the particle 2 is
\be{fG}
F_{1\rightarrow 2}=-\frac{\partial}{\partial l_2} U_{1\rightarrow 2}= 
-\frac{m_1 m_2 G_1^{1/2}G_2^{1/2}}{\left(\sqrt{\frac{G_2}{G_1}}l_1 +l_2\right)^2}.
\ee 
It is directed along the line joining the particles. Note that this force does not satisfy the law of action and reaction. This was to be expected since the external  $G$-wall breaks translation and hence momentum conservation. 

One may perform the same calculation for the Coulomb interaction if the two particles are charged. One then has 
$$
\tilde U= -\frac{\tilde e_1\tilde e_2}{{\tilde r_{12}}} ,
$$
but the units of the charge $e$ are those of $(\mbox{action}\times\mbox{velocity})^{1/2}$, and therefore
$$
\tilde e = e,
$$
unlike $\tilde m = \sqrt{G} m$. Therefore, in the Coulomb case Eq. \eq{potG} is replaced by 
\be{potq}
U_{1\rightarrow 2}= G^{-1/2}_2 \tilde U=   -\frac{\tilde e_1\tilde e_2}{\left(\sqrt{\frac{G_2}{G_1}}l_1 +l_2\right)} \ ,
\ee
while \eq{fG} is replaced by,
\be{fQ}
F_{1\rightarrow 2}=-\frac{\partial}{\partial l_2} U_{1\rightarrow 2}= 
-\frac{e_1e_2}{\left(\sqrt{\frac{G_2}{G_1}}l_1 +l_2\right)^2}.
\ee 

It is important to realize that, as shown by the above Coulomb example, the presence of the $G$-wall affects all the interactions between matter on different sides of the wall and not just the gravitational ones. On the other hand, if one considers experiments realized within a given domain, only the gravitational interaction will be different in different domains. For example, the orbital period of a satellite (in atomic units) in a given elliptic orbit, specified by its eccentricity (dimensionless) and the length of its semimajor axis (given in atomic units), will be proportional to $G^{-1/2}$; whereas for the Coulomb analog there will be no $G$ dependence.
  
\subsection{Reflection and refraction of a particle by a $G$-wall}
\label{refl}

\begin{figure}[h]
\begin{center}
  \includegraphics[width=8cm]{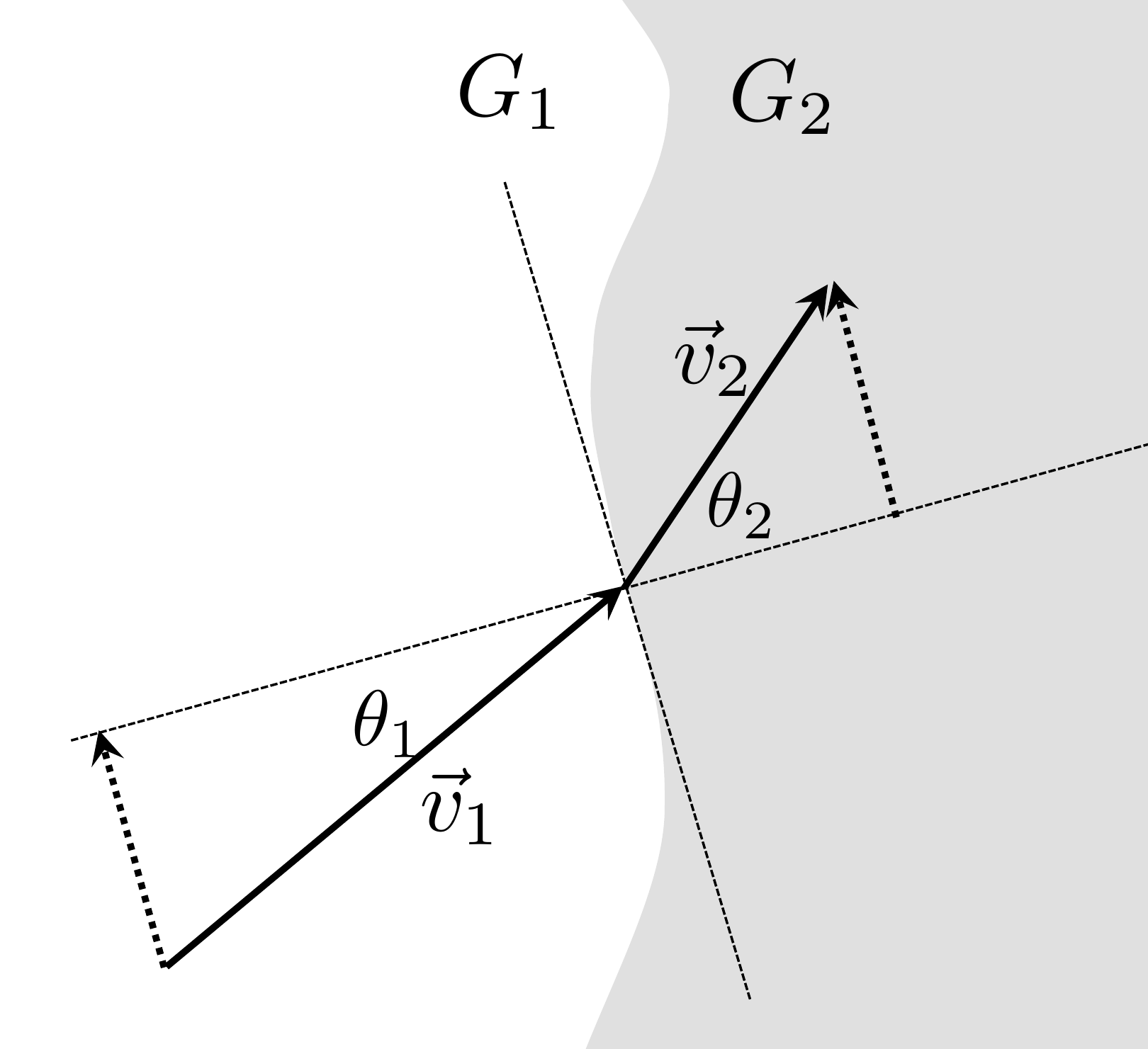}
  \caption{A particle of mass $m$ crosses a $G$-wall dividing a region with gravitational constant $G_1$ from a second region with $G_2$. In gravitational units, the situation is equivalent to that of a free particle that changes its mass as crossing the boundary, from $\tilde m_1=m G_1^{1/2}$ to $\tilde m_2= m G_2^{1/2}$. The speeds in this diagram are for the case in which $G_2>G_1$.}\label{refract}
\end{center}
\end{figure}

We now extend the notion of gravitational units to a relativistic context and examine what happens to a test particle when it crosses a $G$-wall, as it is described in Fig. \ref{refract}. The metric, in gravitational units, $\tg_{\mu\nu}$ is related to  $g_{\mu\nu}$ by
\be{gscale}
g_{\mu\nu}=G \tilde g_{\mu\nu},
\ee
as it follows from the change of coordinates given in Eq. \eq{rescaling}.
We take dimensionless coordinates, so that the metric $g_{\mu\nu}$ has dimensions of length squared, while $\tg_{\mu\nu}$ has units of action. 

We will demand again that the $G$-wall be invisible in terms of gravitational units. That means that $\tg_{\mu\nu}$ should be continuous across the wall, which is equivalent to,
\be{gg}
\frac{1}{G_1}g^{(1)}_{\mu\nu}= \frac{1}{G_2}g^{(2)}_{\mu\nu},
\ee 

implying that {\em the intrinsic geometry of the $G$-wall is different as seen from its two sides}. This novel feature makes $G$-walls essentially different from ordinary domain walls, across which the intrinsic geometry is continuous. 

In term of the tilded variables, the action for the particle to propagate from $1$ to $2$ crossing the $G$-wall may be then written as
\be{action}
I=  -\int  \tilde m(\tilde x) \ \sqrt{-\tg_{\mu\nu}d{\tilde x}^\mu d{\tilde x}^\nu},  
\ee
where
$$
\tilde m(\tilde x)=\left\{\begin{array}{rl}
m\sqrt{G_1}& \mbox{      left of the wall,} \\	
m\sqrt{G_2}& \mbox{      right of the wall.}
\end{array}\right.
$$

Thus the problem is equivalent to a particle whose mass changes for $m\sqrt{G_1}$ to  $m\sqrt{G_2}$ when crossing the $G$-wall. 
The components of the four-momentum parallel to the $G$-wall are not altered (translation invariance in the tangent plane of the worldsheet of the wall, where the effect takes place). Conservation of the spatial momentum implies that the movement occurs in a plane and the Snell-type law,
\be{snell}
v_1\sin\theta_1 = v_2\sin\theta_2.
\ee
On the other hand, conservation of energy yields,    
\be{speed}
\frac{1-v_1^2}{G_1}= \frac{1-v_2^2}{G_2},
\ee
giving the speed of the particle once it emerges to region 2.

When the particle is moving to the region with smaller Newton constant  $G_1>G_2$,  Eq. \eq{speed} indicates that the particle experiments a boost, $v_2^2>v_1^2$. From \eq{snell} we see that in that case $\theta_2<\theta_1$.

In the inverse situation, $G_1<G_2$, the particle will diminish its speed. In that case, Eqs. \eq{snell} and \eq{speed} tell us that, for a given $v_1$, there is a maximal angle for which the particle may cross the wall, 
\be{condtheta}
\sin^2\theta_1 \leq \frac{1}{v_1^2}\left[ 1- \frac{G_2}{G_1}(1-v_1^2)\right].
\ee
If Eq. \eq{condtheta} is not satisfied the particle will experiment a specular bounce off the wall, similar to total internal reflection in geometrical optics. Note, in particular, that if $v_1$ is too small,  the right hand side of the inequality \eq{condtheta} becomes negative, and the particle will bounce for any angle. Also note that in the critical case, when \eq{condtheta} becomes equality,  one has $\theta_2=\pi/2$ and just after crossing the particle moves paralel to the wall.

\section{Action principle}

We will now include the $G$-walls as dynamical objects. This will be achieved by  generalizing the procedure introduced in \cite{Brown:1988kg} to promote the cosmological constant to a dynamical variable that changes across a domain wall possesing a U(1) charge. 

\subsection{Action}

The action takes the form
\be{gaction}
I=I_{grav} + I_{wall} + I_{matter}.
\ee
Here $I_{grav}$, which depends on the metric field $g_{\mu\nu}$,  a three form potential $A_{\mu\nu\rho}$ and a pseudoscalar  $D$ is given by
\be{Igrav}
I_{grav}= \frac{1}{16\pi}\int d^4x \ \sqrt{-\tg}\tilde R -\int (\partial_\alpha D) \A^{\alpha} \ d^4 x.
\ee
The first term in Eq. \eq{Igrav} is the Einstein-Hilbert action for the tilded metric.
The second term is built out of the pseudoscalar vector density $\A^\alpha$, 
$$
\A^\alpha=\frac{1}{3!} \epsilon^{\alpha\beta\gamma\delta}A_{\beta\gamma\delta},
$$
dual to the three-form potential $A_{\alpha\beta\gamma}$ and the field $D$.
One sees from Eq. \eq{Igrav} that $D$ is canonically conjugate of $\A^0=A_{123}$.  

The second term in Eq. \eq{gaction} is an integral over the worldvolume of the $G$-wall,
\be{Gwall}
I_{wall}= \int A - \frac{\tilde\mu}{4\pi}\int\sqrt{-\det{\tilde \gamma}}\  d^3 \sigma \ .
\ee 
Here $\tilde \mu$ is the  tension of the wall in gravitational units, and $\tilde \gamma_{ab}$ is the induced tilded metric (which is the same from both sides of the wall) on it,
\be{induced}
\tilde \gamma_{ab}=\pd{z^\mu}{\sigma^a}\pd{z^\nu}{\sigma^b} \tg_{\mu\nu},
\ee
where the embedding of the wall in spacetime is ${z^\mu = z^\mu (\sigma^a)}$, $a=0,1,2$,  with dimensionless parameters $\sigma^a$. 

The relation between the metric $g_{\mu\nu}$, which is the one defining distances in atomic units (sometimes referred to as the  metric in the ``Jordan frame"), with the tilded metric (sometimes referred to as the metric in the ``Einstein frame") is given by \eq{gscale} but where now $G$ is a function of $D$, $G= G(D)$, which will be chosen below.

The matter action takes the form,
\be{matter}
I_{matter} = \int d^4 x \ \L_m.
\ee
Here the matter Lagrangian density $\L_m$ depends on the matter fields and $g_{\mu\nu}$. It becomes a function of $D$ and $\tg_{\mu\nu}$ through $g_{\mu\nu}=G(D)\tg_{\mu\nu}$.

\subsection{Equations of motion}

Although we take the physical metric as $g_{\mu\nu}$, it is simpler to work with $\tg_{\mu\nu}$ which will be continuous across the wall, because $G$-walls are  isometrically embedded in $\tilde g_{\mu\nu}$, and not in $g_{\mu\nu}$. We will, therefore, vary the action taking $\tg_{\mu\nu}$ as an independent field. 

Varying the action with respect to $A$ we obtain that $D$ is domain-wise constant,
jumping across $G$-walls as
\be{jump}
D_2-D_1 = 1.
\ee
(We have taken in \eq{Gwall} the U(1) charge of the wall to be equal to unity).
  The identification of the two sides with the subscripts ``1'' or ``2" depends on the orientation of the worldsheets.
  
  Varying with respect to $D$ we get 
\be{Feq}
\F\equiv \partial_\alpha\A^\alpha=-\frac{1}{4} T\sqrt{-\tg} \frac{d}{dD} G^2(D),
\ee
where $T$ is the trace of the energy momentum tensor of the matter (defined in terms of the microscopic metric $g_{\mu\nu}$).

The time derivative of  $\A^i$  does not appear in Eq. \eq{Feq}, which means that the evolution of $\A^i$ is arbitrary,  i.e., $\A^i$ is pure gauge. As a consequence, $\partial_i \A^i$ is also arbitrary, and therefore the time evolution of $\A^0$ is arbitrary, and hence it also is pure gauge, with exception of the gauge invariant global mode
\be{global}
P=\int \A^0 \ d^3 x .
\ee
It follows from \eq{Feq} that
$$
\dot P= -\frac{1}{4}\int d^3 x  T\sqrt{-\tg} \frac{d}{dD} G^2(D).
$$
Therefore, after the gauge freedom is factored out, the fields $D,\ A$ have only one physical degree of freedom (not one per point) of ``action-angle'' variables, $D$, $P$. The variable $D$, canonically conjugate to $P$, is domain-wise independent of time (and space).

Note that if $[G^2]''$ and $T$ are different from zero, one may use \eq{Feq} to express $D$ in terms of $\F$ and $T$. It is then permisible to insert that solution into the action \eq{gaction} to obatin a reduced action that involves $A$ but not $D$. This will be the case of interest in the present work. However, the non-invertible case is not devoid of interest; an analog of $G^2(D)=D$, whose second derivative vanishes, was used in \cite{Henneaux:1989zc} to interpret the global mode \eq{global} as a ``cosmic time'', conjugate to the cosmological constant. 

Lastly, varying with respect to the metric field $\tg_{\mu\nu}$ gives rise to
the Einstein equations for this metric,  with $\tilde G=1$. The energy momentum tensor in them has two contributions: one from the $G$-wall, proportional to the tension $\tilde\mu$ and supported on its worldsheet; the other, the energy momentum $\tilde T^{\mu\nu}$ of the matter field obtained by varying $I_m$ with respect to $\tilde g_{\mu\nu}$.  One recovers the equations in atomic units by rescaling with $G(D)$ as in Eq. \eq{gscale}.

\subsection{Interaction of $G$-walls with vacuum energy}

The simplest form of matter one may consider is a constant microscopic vacuum energy density $u$, which gives rise to a cosmological constant 
\be{Lat}
\Lambda = 8\pi Gu.
\ee
 The corresponding matter action is 
\be{action}
I_{matter}=-\int d^4x\ u\sqrt{-g} =
-\int d^4x\ uG^2\sqrt{-\tg}.
\ee
By the very definition of atomic units the constant $u$ is of order unity.

Now  Eq. \eq{Feq} takes the form,
\be{dd}
\F =u\sqrt{-\tilde g}\frac{d}{dD} G(D)^2. 
\ee
We will choose the function $G(D)$ so that: (i) bubble nucleation decreases $G$, making it vanish asymptotically without  ever becoming negative, (ii) no fine tuning is necessary.  

Thus, in this view, the present small value of $G$ is still decreasing,  but at an extremely small rate.  

Such functions $G(D)$ do exist. A simple choice is,
\be{good}
G^2(D) = \frac{G_0^2}{D^2},
\ee
and it is the one we will use bellow.
Note that $G$ {\it decreases} when $D$ {\it increases}. (The possibility of such an ``attractor point" in bubble nucleation has been considered in a different setting and through a different mechanism in Ref. \cite{Dvali:2003br})

\section{Production of $G$-walls by tunneling and thermal activation. Instanton and thermalon} 
\label{production}

In \cite{Brown:1988kg} and  in \cite{Gomberoff:2003zh}, the cosmological constant was relaxed by nucleating membranes due to quantum tunneling (``going through the potential barrier'') and thermal activation (``jumping over the potential barrier'') respectively. The same phenomena occurs here, where nucleation of $G$-walls relaxes the gravitational constant. 



We will now study the probability for nucleating $G$-walls through both processes in the semiclassical approximation. In doing so we will consider only the uniform vacuum energy density $u$ case discussed above, but we will keep the function $G(D)$ generic.

The probability  is of the form,
\be{prob}
P = C\exp(I_E),
\ee
where $I_E$ is the Euclidean action  evaluated on an appropriate extremum,  and $C$ is a slowly varying function.  The Euclidean action is obtained from the exponent $iI$ of the exponential in the Lorentzian path integral, by replacing in it $(\tau,x^0,x^i,A_{0ij},A_{ijk},D)$ by $(-i\tau,-ix^0,x^i,-A_{0ij},iA_{ijk},D)$, and demanding the new variables to be real. This gives
\begin{eqnarray}
I_E &=& \frac{1}{16\pi}\int d^4x \ \sqrt{\tg}\tilde R  +\int  d^4 x(\partial_\alpha D)\A^{\alpha} - \int A  \nonumber \\  &&  - \frac{\tilde\mu}{4\pi}\int  d^3 \sigma\sqrt{\det{\tilde \gamma}} - \int d^4x \ uG^2\sqrt{-\tg} \ , \  \ \ \ \ \ \ \ \ \ \ \  \label{ie}
\end{eqnarray}
which is understood  to be a functional of $\tg_{\mu\nu}$, $A_{\mu\nu\rho}$,$D$, and $z^\mu$.
Note that the action is linear in $A$, and therefore when evaluating it on-shell, the second and third term cancel. The only $D$ dependence left is through  $G(D)$ in the matter term of \eq{ie}.
The probability of nucleating $G$-walls will, therefore, be expressible in terms of $G$ and its change $\Delta G$ due to the nucleation. The change $\Delta G$ in a nucleation will be determined  through the function of $G(D)$ by the  $U(1)$ charge of the bubble, taken here by convention equal to unity, $\Delta D=1$, (Eq. \eq{jump}). 
 
\subsection{Euclidean worldsheet of a spherical $G$-wall}

One expects the extremum relevant for the semiclassical approximation to have the highest possible symmetry. Since we will be interested in including black holes, we will take in the succeeding discussion the symmetry to be just ordinary spherical symmetry SO(3), rather than SO(4). The symmetry will become SO(4) when there is no black hole.

Consider a spherical $G$-wall,  dividing spacetime in two regions, the ``interior'' (``-") and the ``exterior" (``+"), such that the metric on each side is of Schwarzschild-de Sitter form, 
\be{metric}
d\tilde s^2_\pm = \tilde f^2_{\pm}d\tilde t_{\pm}^2 + \tilde f^{-2}_{\pm} d\tilde r^2 + \tilde r^2d\Omega^2, 
\ee 
\be{f2}
\tilde f_{\pm}^2= 1-\frac{2\tilde M_{\pm}}{\tilde r}-\frac{\tilde r^2}{\tilde l^2_{\pm}}.
\ee
(In order to make direct contact with previous results \cite{Gomberoff:2003zh}, we use in this subsection time and radial coordinates with dimensions of length).

In \eq{f2},
\be{l}
\frac{1}{\tilde l_{\pm}^2}=\frac{\tilde\Lambda_{\pm}}{3} =\frac{8\pi}{3}\tilde u_{\pm}=
 \frac{8\pi}{3}G_{\pm}^2u,
\ee
and the coordinate $\tilde t_\pm$ is defined such that increases anticlockwise around the cosmological horizon.
The wall is parameterized by,
\be{para}
\tilde r= R(\tau), \ \ \ \tilde t_{\pm}=T_{\pm}(\tau), 
\ee
with $\tau$ being the proper time,
\be{proper}
\tilde f_{\pm}^2 \dot T_{\pm}^2 + \tilde f_{\pm}^{-2} \dot R^2 = 1.
\ee

The matching in the geometries at the joining membrane  gives a first integral for the equation of motion of $R(\tau)$,
\be{df2}
\sqrt{\tilde f^2_- -\dot { R}^2}- \sqrt{\tilde f^2_+ -\dot {R}^2} = \sigma\tilde \mu R .
\ee
Here $\sigma=1$, when the cosmological horizon is in the exterior, which will be the case for the instanton discussed below; and $\sigma=-1$ when the cosmological horizon is in the interior, which will be the case for the thermalon discussed below.  (See \cite{Gomberoff:2003zh} for details). 
\subsection{Radius}

There are two types of solutions of Eq.\eq{df2} that are of interest here. The instanton described in \cite{Brown:1988kg} for the nucleation of membranes by tunneling; and the thermalon, a static solution found in \cite{Gomberoff:2003zh} for the nucleation by thermal activation. In both cases it is necessary to determine the radius of formation $R=\rho$ which occurs when $\dot R=0$. From \eq{df2}, at that instant,
\be{cth}
\Delta \tilde M \equiv \tilde M_- - \tilde M_+ = \frac{1}{2}(\alpha^2-\tilde\mu^2){\rho}^3 - 
 \sigma\tilde\mu \tilde f_+
{\rho}^2,
\ee
where 
\be{alpha2}
\alpha^2 =\frac{8}{3}\pi u (G^2_+ -G^2_-).
\ee

For the instanton  $\sigma=1$ in Eq. \eq{df2}. Here there is no black hole, hence $\tilde M_{\pm}=0$, and the geometry is that of de Sitter on each side of the $G$-wall.  In this case Eq. \eq{cth} is quadratic, and it has a solution if and only if $\alpha^2>\tilde\mu^2>0$, that is, when
\be{GG}
G_+^2>G_-^2,
\ee
so that the notation in \eq{alpha2} is justified.
The radius of formation is given by
\be{rhoinst}
\rho=2\tilde\mu\left([\alpha^2-\tilde\mu^2]^2 + \frac{4{\tilde\mu}^2}{\tilde l_+^2} \right)^{-1/2}.
\ee

The function $R(\tau)$ is a ``bounce" going from $R=0$ to a maximum $R=\rho$ and back to $R=0$. The Euclidean $G$-wall worldsheet is a three-sphere of radius $\rho$ where the two de Sitter geometries are glued together.  
 Once the wall materializes its radius increases without limit (see \cite{Brown:1988kg} for details).

For the thermalon $\Delta \tilde M$ does not vanish. Its value is determined by demanding the right hand side of Eq.  \eq{cth} to be an extremum in $\rho$, which is equivalent to requiring $\ddot R=0$. Once this value is inserted on the left hand side of  \eq{cth} the resulting equation cannot be solved analytically, but can be analyzed graphically.  One may show that solutions with  positive $\alpha^2$ also exist. That analysis is given in \cite{Gomberoff:2003zh} and  will be not repeated here.  

 The thermalon is static, therefore $\dot R=0$, and the $G$-wall stays always at $R=\rho$ unless it is perturbed. 
  Of particular interest is the ``cosmological thermalon" in thermal equilibrium with the cosmological horizon. For that solution the cosmological horizon is in the interior so that $\sigma=-1$ in \eq{df2}.  After nucleation, if slightly perturbed with an inward velocity, the $G$-wall will collapse to a black hole, even if originally there is none ($M_+=0$, $M_-\neq 0$).

\subsection{Probability}

It is simpler to evaluate the on-shell action in the Hamiltonian formalism. This is so because: (i) within each of the domains separated by the wall one may use a time coordinate in which the metric is static, therefore the $\pi^{ij}\dot g_{ij}$ term vanishes. (ii) The total ``bulk'' Hamiltonian, which is a linear combination of the constraints vanishes, because on-shell the constraints hold.(iii)  When a horizon is present in the interior (final region) one just adds one fourth of its area, $\tilde{\cal A}_{-}$ in the tilded metric\cite{Banados:1993qp}. Therefore  the on-shell action, reduces to the $p \dot q$ term of the $G$-wall, which, rewritten in term of the velocities is given by,
\be{evaction}
I_E=\frac{1}{4}\tilde{\mathcal A}_{-}  -\int A - \frac{\tilde\mu}{4\pi}\int d^3 \sigma\sqrt{\det{\tilde \gamma}}.
\ee

The second term in this expression is ill defined, because the value of $A$ jumps as one crosses the $G$-wall. The regularized value of the integral must be taken to be,
\be{regint} 
\int A =  \F_{on} \ V^{(4)}_- ,
\ee
where $V^{(4)}_-$ is the volume of the interior region and $\F_{on}$ is the value of $\F$ on the membrane, which may be defined by thickening the membrane and using the mean value theorem as 
\be{Fon}
\F_{on}=\left(\frac{1}{D_+-D_-}\right)\int_{D_-}^{D_+} \F  \ dD
.
\ee
Eq. \eq{regint} may be justified by realizing that the variation of its right hand side with respect to the membrane coordinates should give the ``electric force'' $q\F_{on}$, where ${q=D_+-D_-=1}$. On the other hand, the Euclidean value of $\F$ is given by the counterpart of \eq{dd},
\be{dd}
\F =-u\sqrt{\tilde g}\ \frac{d}{dD} G(D)^2. 
\ee
 Inserting this in \eq{Fon}, and carrying out the integration, yields 
\be{AG1}
\int A = uV^{(4)}_- (G_-^2 - G_+^2).
\ee

The final regularized action is then
\be{AG}
I_E = \frac{1}{4}(\tilde{\mathcal A}_{-} -\tilde{\mathcal A}_{+}) - \frac{\tilde\mu}{4\pi}V^{(3)} - uV^{(4)}_- (G_-^2 - G_+^2),
\ee
where $V^{(3)}$ is the three-volume of the history of the $G$-wall, and $\tilde{\mathcal A}_{+}$ is the area of the horizon that the interior region would have if the $G$-wall were absent.
As explained in Sec. IV of \cite{Gomberoff:2003zh}, the substractions of $G_+^2$  and  of $\tilde{\mathcal A}_{+}/4$ in \eq{AG}, embody the  requirement that the fields in the exterior should be those corresponding to the solution of the equations of motion that would hold everywhere if the transition where never to occur, which ensures that the probability of creating $G$-walls with $\mu=u=0$ is equal to unity.  
 
 Eq. \eq{AG} is the same that was obtained in Ref. \cite{Gomberoff:2003zh} in the context of the cosmological constant problem. Therefore all the analysis carried out there for the probability in various limiting cases can be taken over simply by substituting Eq. \eq{alpha2} for $\alpha^2$ in that reference. 
  
Instead we would like to devote some space to bring out how the formula for the probability captures in a nutshell the fact that, in the case of G, the decay process relates two different scales, a feature which has no analog in the cosmological constant case. This may be seen in the simplest manner for the instanton in the approximation in which the nucleation radius $\rho$ is much smaller that the dS radius $\tilde l_+$. In that case we may approximate the interior volume by its flat space expression, ${V_-^{(4)} = \frac{1}{2}\pi^2 \rho^4}$, whereas ${V_3=2\pi^2\rho^3}$. Therefore the action takes the form
\be{instapp}
I_E^{\tiny instanton}=-\frac{\tilde\mu\pi}{2} \rho^3 + \frac{3\pi}{16}\alpha^2\rho^4.
\ee
Minimizing this action with respect to $\rho$ we obtain the radius of formation,
$$
\rho=\frac{2\tilde\mu}{\alpha^2},
$$
which could also be obtained directly from Eq. \eqref{rhoinst}. Inserting this in \eqref{instapp} gives,
\be{instappac}
I_E^{\tiny instanton}=-\frac{\pi \tilde\mu^4}{\alpha^6}.
\ee
The probability of formation is therefore given by,
\begin{eqnarray}
P &\sim &  
 \exp{\left(\frac{1}{\hbar}I_E^{\tiny instanton}\right)}=
 \nonumber \\
 &=&  \exp\left[-\left(\frac{3}{8}\right)^3 \frac{\tilde \mu^4}{\pi^2 \hbar u^3(G^2_+ - G^2_-)^3}\right],\label{P4}
\end{eqnarray}
where we have substituted for $\alpha^2$ in \eq{instappac} its value \eq{alpha2}.
If $D>>1$, which begins to be true after a few steps, and becomes more and more valid throughout the present time, then $G_+\approx G_- =G$, and we may write this,
\be{probab}
P \sim  \exp\left(-\left[\frac{27\pi^2}{8^4|(\log G)'|}\right]\left[\frac{\tilde \mu^4}{\hbar u^3G^6}\right]\right).
\ee
 The first factor in the exponent is of order unity, whereas the second is more illuminating. By using $\tilde u = G^2 u$, $\tilde \mu = G^{3/2} \mu$, it maybe rewritten as, 
\be{talk}
  \frac{\mu^4}{\hbar u^3}= \frac{\tilde \mu^4}{\hbar \tilde u^3}.
\ee
Each of the two sides of the equality contains a term which is natural (i.e., of order unity) in atomic units and another which is natural in gravitational units. For example, on the left hand side, $\mu$ is of gravitational nature because it is the tension of a $G$-wall, so its tilded value is always of order unity but its untilded value depends on the epoch: it is of order unity at the beginning, but very large at present. On the other hand $u$, which is of atomic nature is always of order unity. Thus, the probability \eq{probab} is of order unity at the beginning and very small at present. The same conclusion is of course obtained looking at the right hand side.  Now $\tilde \mu$ is always of order unity, but $\tilde u$ depends on the epoch.

\section{The gravitational constant as a thermodynamic black hole parameter}

When a black hole metric is expressed in terms of the mass, angular momentum, and other charges, it depends explicitly, in addition, on the gravitational constant $G$ and the cosmological constant $\Lambda$. The standard black hole thermodynamics theorems (see, for example, \cite{hawking1973large}) assume that both $G$ and $\Lambda$ are universal constants; therefore the question of whether the black hole entropy can only increase in a irreversible process has to be re-analyzed when $G$ and $\Lambda$ are allowed to vary. For the cosmological constant, this analysis was performed in the simple case of spherical symmetry in Ref. \cite{Teitelboim:1985dp} and the question was answered in the affirmative. It is the purpose of this section to address the question for $G$. 

The issue at hand becomes already manifest for the case of a Schwarzschild black hole,
\be{Sch}
ds^2=- \left(1-\frac{2MG}{r}\right)dt^2 +\left(1-\frac{2MG}{r}\right)^{-1} dr^2
+r^2d\Omega_2^2 \ ,
\ee
for which the entropy is given by 
\be{Sbh}
S=\frac{4\pi G M^2}{\hbar}.
\ee
When the hole absorbs a $G$-wall both $G$ and $M$ will change. It may happen that the value of $G$ diminishes as it indeed is the case, for example, for the thermalon discussed in subsection IV.B (recall Eq. \eq{GG}). Now, if we had coupled the $G$-wall to $g_{\mu\nu}$ as a standard domain wall,\footnote{For a standard domain wall coupling, the Hilbert term in the action \eq{gaction}, $\sqrt{-\tg}\tilde R$, would be replaced by
$G(D)^{-1}\sqrt{-g} R$.
No $\tilde g_{\mu\nu}$ would be introduced, the factor $G(D)^{-1}$ would remain ``outside of the curvature''. Straightforward analogs of the action so obtained may be written to make dynamical, for example,  the cosmological constant, any parameter appearing in a field theory Lagrangian (coupling constants, masses) or the string tension. In none of these cases there is a problem such as the one that arises with a dynamical $G$ in connection with black hole thermodynamics. 

 For the case of the cosmological constant, if one sets ${\Lambda(D)=4\pi G e^2 D^2 +\lambda}$ in the Hilbert Lagrangian, one obtains the action employed in \cite{Brown:1988kg}. (Here $e$ is the charge of the standard domain wall and $\lambda$ a ``bare cosmological constant").}  the mass $M$ of the hole would increase after absorption of the wall, since the latter would have positive energy; but there would be no guarantee that the increasing $M$ would be sufficient to overcompensate the decreasing $G$ so as to make the entropy increase, since both changes are independent. Thus, the second law of black hole thermodynamics could be easily violated and one would face a major difficulty.

The way out is provided by realizing that in what we have called gravitational units the entropy  \eq{Sbh} reads,
\be{Sbht}
S=\frac{4\pi {\tilde M}^2}{\hbar},
\ee
and therefore if one couples the domain wall in the {\it standard manner in gravitational units}, as in \eq{gaction}, the problem is solved because now $\tilde M$ increases and so does the entropy.  The argument holds as well for a general black hole, because in gravitational units, the standard theorems apply. 

The value of $D$, which determines $G$, may be thought of as a charge, which counts - up to an additive constant - the number of $G$-walls absorbed by the hole. Its conjugate chemical potential is the integral of $A_{0\theta\phi}$ at infinity (or at the cosmological horizon), when one demands $A_{0\theta\phi}$ to vanish at the black hole horizon.

It is quite remarkable that the second law of black hole thermodynamics should dictate how a dynamical $G$ must be incorporated in the action principle if one wants to relate the cosmological and microscopical scales. But perhaps it should not be surprising, since after all
$$
S=\frac{c^3}{4\hbar G}(\mbox{area of black hole horizon}),
$$ 
is {\it the one} formula where those two scales appear explicitly rather than just being related through dimensional analysis.

\section*{Acknowledgments}

The Centro de Estudios Cient\'ificos (CECs) is funded by the Chilean Government
through the Centers of Excellence Base Financing Program of Conicyt. C.B. wishes to thank the Alexander von Humboldt Foundation for a Humboldt Research Award.
The work of A.G. was partially supported by Fondecyt (Chile) Grant $\#$1141309. The authors express their gratitude to the Max Plank Institute for Gravitational Physics at Golm, where part of this work was carried out.  They benefited from illuminating comments by G.~Dvali, G.~Gabadadze, M.~Porrati, and other colleagues at the Center for Cosmology and Particle Physics at New York University. 

%


\begin{thebibliography}{11}%
\makeatletter
\providecommand \@ifxundefined [1]{%
 \@ifx{#1\undefined}
}%
\providecommand \@ifnum [1]{%
 \ifnum #1\expandafter \@firstoftwo
 \else \expandafter \@secondoftwo
 \fi
}%
\providecommand \@ifx [1]{%
 \ifx #1\expandafter \@firstoftwo
 \else \expandafter \@secondoftwo
 \fi
}%
\providecommand \natexlab [1]{#1}%
\providecommand \enquote  [1]{``#1''}%
\providecommand \bibnamefont  [1]{#1}%
\providecommand \bibfnamefont [1]{#1}%
\providecommand \citenamefont [1]{#1}%
\providecommand \href@noop [0]{\@secondoftwo}%
\providecommand \href [0]{\begingroup \@sanitize@url \@href}%
\providecommand \@href[1]{\@@startlink{#1}\@@href}%
\providecommand \@@href[1]{\endgroup#1\@@endlink}%
\providecommand \@sanitize@url [0]{\catcode `\\12\catcode `\$12\catcode
  `\&12\catcode `\#12\catcode `\^12\catcode `\_12\catcode `\%12\relax}%
\providecommand \@@startlink[1]{}%
\providecommand \@@endlink[0]{}%
\providecommand \url  [0]{\begingroup\@sanitize@url \@url }%
\providecommand \@url [1]{\endgroup\@href {#1}{\urlprefix }}%
\providecommand \urlprefix  [0]{URL }%
\providecommand \Eprint [0]{\href }%
\providecommand \doibase [0]{http://dx.doi.org/}%
\providecommand \selectlanguage [0]{\@gobble}%
\providecommand \bibinfo  [0]{\@secondoftwo}%
\providecommand \bibfield  [0]{\@secondoftwo}%
\providecommand \translation [1]{[#1]}%
\providecommand \BibitemOpen [0]{}%
\providecommand \bibitemStop [0]{}%
\providecommand \bibitemNoStop [0]{.\EOS\space}%
\providecommand \EOS [0]{\spacefactor3000\relax}%
\providecommand \BibitemShut  [1]{\csname bibitem#1\endcsname}%
\let\auto@bib@innerbib\@empty
\bibitem [{\citenamefont {Dirac}(1937)}]{Dirac:1937ti}%
  \BibitemOpen
  \bibfield  {author} {\bibinfo {author} {\bibfnamefont {P.~A.~M.}\
  \bibnamefont {Dirac}},\ }\href {\doibase 10.1038/139323a0} {\bibfield
  {journal} {\bibinfo  {journal} {Nature}\ }\textbf {\bibinfo {volume} {139}},\
  \bibinfo {pages} {323} (\bibinfo {year} {1937})}\BibitemShut {NoStop}%
\bibitem [{\citenamefont {Brown}\ and\ \citenamefont
  {Teitelboim}(1988)}]{Brown:1988kg}%
  \BibitemOpen
  \bibfield  {author} {\bibinfo {author} {\bibfnamefont {J.~D.}\ \bibnamefont
  {Brown}}\ and\ \bibinfo {author} {\bibfnamefont {C.}~\bibnamefont
  {Teitelboim}},\ }\href {\doibase 10.1016/0550-3213(88)90559-7} {\bibfield
  {journal} {\bibinfo  {journal} {Nucl.Phys.}\ }\textbf {\bibinfo {volume}
  {B297}},\ \bibinfo {pages} {787} (\bibinfo {year} {1988})}\BibitemShut
  {NoStop}%
\bibitem [{\citenamefont {Henneaux}\ and\ \citenamefont
  {Teitelboim}(1989)}]{Henneaux:1989zc}%
  \BibitemOpen
  \bibfield  {author} {\bibinfo {author} {\bibfnamefont {M.}~\bibnamefont
  {Henneaux}}\ and\ \bibinfo {author} {\bibfnamefont {C.}~\bibnamefont
  {Teitelboim}},\ }\href {\doibase 10.1016/0370-2693(89)91251-3} {\bibfield
  {journal} {\bibinfo  {journal} {Phys. Lett.}\ }\textbf {\bibinfo {volume}
  {B222}},\ \bibinfo {pages} {195} (\bibinfo {year} {1989})}\BibitemShut
  {NoStop}%
\bibitem [{\citenamefont {Gomberoff}\ \emph {et~al.}(2004)\citenamefont
  {Gomberoff}, \citenamefont {Henneaux}, \citenamefont {Teitelboim},\ and\
  \citenamefont {Wilczek}}]{Gomberoff:2003zh}%
  \BibitemOpen
  \bibfield  {author} {\bibinfo {author} {\bibfnamefont {A.}~\bibnamefont
  {Gomberoff}}, \bibinfo {author} {\bibfnamefont {M.}~\bibnamefont {Henneaux}},
  \bibinfo {author} {\bibfnamefont {C.}~\bibnamefont {Teitelboim}}, \ and\
  \bibinfo {author} {\bibfnamefont {F.}~\bibnamefont {Wilczek}},\ }\href
  {\doibase 10.1103/PhysRevD.69.083520} {\bibfield  {journal} {\bibinfo
  {journal} {Phys.Rev.}\ }\textbf {\bibinfo {volume} {D69}},\ \bibinfo {pages}
  {083520} (\bibinfo {year} {2004})}  \BibitemShut {NoStop}%
\bibitem [{\citenamefont {Jordan}(1949)}]{Jordan:1949zz}%
  \BibitemOpen
  \bibfield  {author} {\bibinfo {author} {\bibfnamefont {P.}~\bibnamefont
  {Jordan}},\ }\href {\doibase 10.1038/164637a0} {\bibfield  {journal}
  {\bibinfo  {journal} {Nature}\ }\textbf {\bibinfo {volume} {164}},\ \bibinfo
  {pages} {637} (\bibinfo {year} {1949})}\BibitemShut {NoStop}%
\bibitem [{\citenamefont {Brans}\ and\ \citenamefont
  {Dicke}(1961)}]{PhysRev.124.925}%
  \BibitemOpen
  \bibfield  {author} {\bibinfo {author} {\bibfnamefont {C.}~\bibnamefont
  {Brans}}\ and\ \bibinfo {author} {\bibfnamefont {R.~H.}\ \bibnamefont
  {Dicke}},\ }\href {\doibase 10.1103/PhysRev.124.925} {\bibfield  {journal}
  {\bibinfo  {journal} {Phys. Rev.}\ }\textbf {\bibinfo {volume} {124}},\
  \bibinfo {pages} {925} (\bibinfo {year} {1961})}\BibitemShut {NoStop}%
\bibitem [{\citenamefont {Dicke}(1962)}]{PhysRev.125.2163}%
  \BibitemOpen
  \bibfield  {author} {\bibinfo {author} {\bibfnamefont {R.~H.}\ \bibnamefont
  {Dicke}},\ }\href {\doibase 10.1103/PhysRev.125.2163} {\bibfield  {journal}
  {\bibinfo  {journal} {Phys. Rev.}\ }\textbf {\bibinfo {volume} {125}},\
  \bibinfo {pages} {2163} (\bibinfo {year} {1962})}\BibitemShut {NoStop}%
\bibitem [{\citenamefont {Dvali}\ and\ \citenamefont
  {Vilenkin}(2004)}]{Dvali:2003br}%
  \BibitemOpen
  \bibfield  {author} {\bibinfo {author} {\bibfnamefont {G.}~\bibnamefont
  {Dvali}}\ and\ \bibinfo {author} {\bibfnamefont {A.}~\bibnamefont
  {Vilenkin}},\ }\href {\doibase 10.1103/PhysRevD.70.063501} {\bibfield
  {journal} {\bibinfo  {journal} {Phys. Rev.}\ }\textbf {\bibinfo {volume}
  {D70}},\ \bibinfo {pages} {063501} (\bibinfo {year} {2004})}  \BibitemShut {NoStop}%
\bibitem [{\citenamefont {Banados}\ \emph {et~al.}(1994)\citenamefont
  {Banados}, \citenamefont {Teitelboim},\ and\ \citenamefont
  {Zanelli}}]{Banados:1993qp}%
  \BibitemOpen
  \bibfield  {author} {\bibinfo {author} {\bibfnamefont {M.}~\bibnamefont
  {Banados}}, \bibinfo {author} {\bibfnamefont {C.}~\bibnamefont {Teitelboim}},
  \ and\ \bibinfo {author} {\bibfnamefont {J.}~\bibnamefont {Zanelli}},\ }\href
  {\doibase 10.1103/PhysRevLett.72.957} {\bibfield  {journal} {\bibinfo
  {journal} {Phys. Rev. Lett.}\ }\textbf {\bibinfo {volume} {72}},\ \bibinfo
  {pages} {957} (\bibinfo {year} {1994})}  \BibitemShut {NoStop}%
\bibitem [{\citenamefont {Hawking}\ and\ \citenamefont
  {Ellis}(1973)}]{hawking1973large}%
  \BibitemOpen
  \bibfield  {author} {\bibinfo {author} {\bibfnamefont {S.}~\bibnamefont
  {Hawking}}\ and\ \bibinfo {author} {\bibfnamefont {G.}~\bibnamefont
  {Ellis}},\ }\href {https://books.google.cl/books?id=QagG\_KI7Ll8C} {\emph
  {\bibinfo {title} {The Large Scale Structure of Space-Time}}},\ Cambridge
  Monographs on Mathematical Physics\ (\bibinfo  {publisher} {Cambridge
  University Press},\ \bibinfo {year} {1973})\BibitemShut {NoStop}%
\bibitem [{\citenamefont {Teitelboim}(1985)}]{Teitelboim:1985dp}%
  \BibitemOpen
  \bibfield  {author} {\bibinfo {author} {\bibfnamefont {C.}~\bibnamefont
  {Teitelboim}},\ }\href {\doibase 10.1016/0370-2693(85)91186-4} {\bibfield
  {journal} {\bibinfo  {journal} {Phys. Lett.}\ }\textbf {\bibinfo {volume}
  {B158}},\ \bibinfo {pages} {293} (\bibinfo {year} {1985})}\BibitemShut
  {NoStop}%
\end{thebibliography}
\end{document}